\documentclass[preprints,article,accept,moreauthors,pdftex]{Definitions/mdpi} 

\firstpage{1} 
\pubvolume{1}
\issuenum{1}
\articlenumber{0}
\pubyear{2021}
\copyrightyear{2020}
\datereceived{} 
\dateaccepted{} 
\datepublished{} 
\hreflink{https://doi.org/} % If needed use \linebreak
\usepackage{bm}

\Title{High harmonic generation in monolayer and bilayer of transition metal dichalcogenide}

\TitleCitation{Title}

\Author{Yeon Lee $^{1}$, Dasol Kim $^{1}$, Dong Eon Kim $^{1,*}$ and Alexis Chac\'on $^{1}$}

\AuthorNames{Firstname Lastname, Firstname Lastname and Firstname Lastname}

\AuthorCitation{Lastname, F.; Lastname, F.; Lastname, F.}
\address{%
$^{1}$ \quad Department of Physics and Center for Attosecond Science and Technology, POSTECH, 7 Pohang 37673, South Korea; Max Planck POSTECH/KOREA Research Initiative, Pohang, 37673, South Korea; dldusigo@gmail.com (Y. Lee); dsestil@gmail.com (D. Kim); achacon@postech.ac.kr (A. Chac\'on)\\
}

\corres{Correspondence: kimd@postech.ac.kr (D. E. Kim)}  

\abstract{In transition metal dichalcogenides (TMDCs), charge carriers have spin, pseudospin, and valley degrees of freedom associated with magnetic moments. The monolayers and bilayers of the TMDCs, in particular, MoS$_2$, lead strong couplings between the spin and pseudospin effects. This feature have drawn attention to TMDCs for their potential use in advanced tech devices. Meanwhile, high-order harmonic generation (HHG) has recently been applied to the characterization of the electronic structure of solids, such as energy dispersion, Berry-curvature, and topological properties. Here, we show theoretical results obtained with the `philosophy' of using HHG to investigate the structural effects of the monolayer and bilayers of MoS$_2$ on nonlinear optical emission. We use a simple model for MoS$_2$ in the 3R AB staking. 
We find that the pseudospin and valley indexes (the Berry curvature and the dipole transition matrix element) in TMDC driven by circularly polarized laser (CPL) can encode in the high energy photon emissions. This theoretical investigation is expected to pave the way for the ultrafast manipulation of valleytronics and lead to new questions concerning the spin-obit-coupling (SOC) effects on TMDC materials, Weyl Semimetals, and topological phases and transitions in topological insulators. }

\keyword{high-order harmonic generation; transition metal dichalcogenides; pseudospin; Berry curvature; dipole transition matrix element; tight-binding model} 

\begin{document}
%%%%%%%%%%%%%%%%%%%%%%%%%%%%%%%%%%%%%%%%%%
%The order of the section titles is: Introduction, Materials and Methods, Results, Discussion, Conclusions for these journals: aerospace,algorithms,antibodies,antioxidants,atmosphere,axioms,biomedicines,carbon,crystals,designs,diagnostics,environments,fermentation,fluids,forests,fractalfract,informatics,information,inventions,jfmk,jrfm,lubricants,neonatalscreening,neuroglia,particles,pharmaceutics,polymers,processes,technologies,viruses,vision

\section{Introduction}
Since high-order harmonics were first experimentally observed in ZnO~\cite{GhimireNatPhy2011,Liu2017,VampaNat2015}, their potential in the nonlinear optical spectroscopy of solids has attracted the attention of ultrafast sciences and condensed matter physics~\cite{VampaNat2015,Itatani2004,ShambhuNatPhy2019,Symphony2019}. 

\begin{figure*}[!ht]
\begin{center}
\includegraphics[width=14cm]{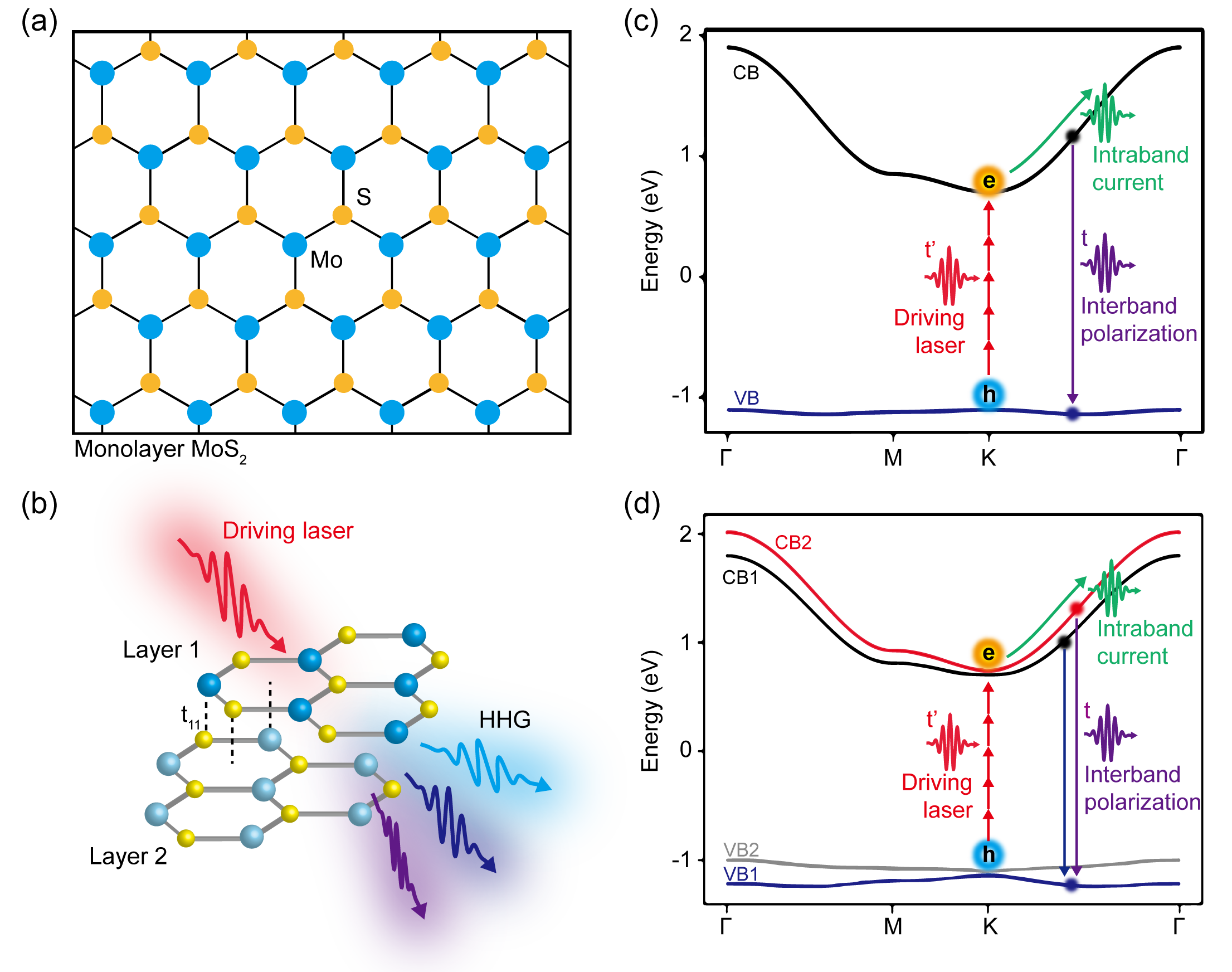}
\caption{Monolayer and 3R-polytype bilayer MoS$_2$. (a) Top view of the monolayer MoS$_2$ (Mo atom is in blue and and Se atoms in yellow, respectively). 3D view of bilayer MoS$_2$ with a MIR driving laser (red-oscillations) and harmonic emission (white-blue, blue and violet osciallations) is shown in (b). The red pulse indicates driving laser field while the blue-purple pulses represent the high harmonic emission. The black vertical dashed lines point out the interlayer coupling between two layers of MoS$_2$ denoted by $t_{11}$. (c) and (d) Schematic diagram of high-order harmonic generation process from the monolayer and bilayer MoS$_2$, respectively. For the monolayer valence and conduction bands,  we choose the tight-binding parameters: $t_1=0.3998$, $t_2=0.066$~{\rm eV} are the NN and NNN parameters (See text) and the `Dirac Mass' or local potential $M_0=0.900$~eV at `Mo' and `S' atoms, described by the laser-field free Hamiltonian of Eq.~(\ref{eq:freeham0}). These parameters match the minimum ($\epsilon_{g}^{(0)}\sim1.8$~eV) and maximum energy between the valence and conduction bands. The electron oscillation in the conduction band generate intraband current while the electron-hole recombination cause the interband currents depicted by vertical blue and violet lines, respectively. }
\label{fig:figure1}
\end{center}
\end{figure*}

The process of High-order Harmonic Generation (HHG) is a highly nonlinear phenomenon in the sense that an incoming middle infrared or infrared (MIR or IR) $\hbar\omega_0\sim 0.309$~eV (with wavelength $\lambda_0=4~\mu{m}$) interacting with the solid produces a new spectrum, as a consequence of the optical responses of the lattice to the strong laser field (illustrated in Fig.~\ref{fig:figure1})~\cite{VampaPRL2014,EOsikaPRX2017,Symphony2019}. The outgoing emission contain high energy photons, $n\hbar\omega_0$, regarding the fundamental driving laser, $\hbar\omega_0$, where $n$ is typically an integer~\cite{CorkumAndFerenkROP2018,EOsikaPRX2017,Symphony2019} (for a mathematically formal description, See Appendixes~ \ref{sec:AppendixA1Bands}~and~\ref{sec:KeldyshSFA}). This paper explores the high-order harmonics emission from monolayer and bilayer of a typical TMDC such as~MoS$_2$.

\noindent The monolayer and bilayer crystalline structures of MoS$_2$ are illustrated in Figs.~\ref{fig:figure1}(a) and~\ref{fig:figure1}(b), respectively ~\cite{AndorPRB2018,NatComYaoWang2013,XuechaoPhysRevB2020}. In the case of the bilayer, the stacking is featured by the AB type stacking which is so called 3R polytype~\cite{XuechaoPhysRevB2020}. This means that the `S1' atom of Layer1 interacts with an `empty space' (dashed black link in Fig.~\ref{fig:figure1}(b)) of Layer2; and, the `Mo' atom of Layer1 interacts with an orbital of `S2' atom in Layer2. This is the nearest neighbour interaction between inter-layers. The interactions between the `Mo' and `S' atoms in different layers are mediated by the hopping parameter $t_{11}$~\cite{AndorPRB2018,NatComYaoWang2013} which is described in Section~\ref{sec:CTDDM0}~and~\ref{MonoAndBiLayerHam0}. 
\noindent The manipulation of monolayers, bilayer up to bulk TMDCs has been studied by several experimental and theoretical efforts using Angular Resolved Photoelectron-Energy Spectrum (ARPES) techniques~\cite{RevModPhys.75.473,ARPESExp2016}. However, its complete characterization remains yet a downright challenging research field. %Recently works by Mrudul $\&$ Dixit~\cite{Dixit2021} have shown interesting dependence of high-order harmonic generation (HHG) spectrum of bi-layers in comparison to monolayers in graphene.  

\noindent Returning to the key physical insight of HHG in solids, the simplified description of the HHG mechanism is depicted in Figs.~\ref{fig:figure1}(c) and~\ref{fig:figure1}(d) for both monolayer and bilayer MoS$_2$~\cite{NatComYaoWang2013}. The accelerated carriers in the conduction (valence) band have two intrinsic mechanisms~\cite{VampaPRL2014}: {\it intraband current} and  {\it interband current}. Note that our theory is subjected to the Keldysh approximation in which $\epsilon_0\geq\hbar\omega_0$, where $\epsilon_0$ is the band gap of the crystal (See Appendix~\ref{sec:KeldyshSFA}). Thus, the interband mechanism of HHG can be interpreted in terms of three simple steps~\cite{VampaJPB2017,Luu2018,VampaPRL2015}: ({\it i}) at a time $t'$ when the driving laser-field, ${\bf E}(t)$, reaches its maximum, an electron-hole ({\it e}) and ({\it h}) pairs is created (see Figs.~\ref{fig:figure1}(c)-\ref{fig:figure1}(d)) by the action of the oscillating external electric field ${\bf E}(t)$, ({\it ii}) between $t'$ and $t$, the {\it e} and {\it h} are accelerated or propagated under the quasi-classical action $S({\bf k},t',t)$~Eq.~(\ref{eqn:ActionML}) in the bands of the MoS$_2$ (See Appendixes~\ref{sec:AppendixA1Bands}~and~\ref{sec:SPASBE}), gaining a `considerable' amount of energy from the laser and bands, and simultaneously encoding rich information of the lattice structure in the crystal (see Fig.~\ref{fig:figure1}(a) and~Fig.~\ref{fig:figure1}(b))~\cite{VampaJPB2017,VampaPRL2014,VampaNat2015,Luu2018,AlexisPRB2020}. `Finally', ({\it iii}) after these {\it e} and {\it h} are accelerated on the Brillouin zone (BZ), the {\it e} and {\it h} can find a time $t$ at which they annihilate each other. Through the conservation of energy~\cite{VampaJPB2017}, this physical picture converts the accumulated {\it e}-{\it h} energy into a high energy photon with respect to the fundamental-laser as depicted in Figs.~\ref{fig:figure1}(c) and~\ref{fig:figure1}(d).

\noindent Surprisingly, this HHG process encodes rich information about the atomic structural, electron dynamics in the TMDCs. This opens the door to investigate fundamental questions and technological development such as symmetries of the Hamiltonian in the solids, the fermion-dynamics, the effects of the Berry curvature, pseudospin, spin-valley, topological phases and transitions~\cite{AlexisPRB2020,DenitsaPRA2021,Luu2018,NicolasNatComm2017,DenitsaACSNano2021,Hasan2010,HaldaneNobel2016}. Additionally, these features are relevant to the study of the fundamental protections of the symmetries, carrier measurements, Bloch oscillations at the natural time-scale of the electrons, sub-femtoseconds ($10^{-18}$~{\it s})~\cite{LangerNature2018}. 

\noindent In addition, it has been demonstrated that in TMDCs, the number of layers and the polytype geometrical features modify the electron structure of the TMDCs and their transport features~\cite{AndorPRB2018,NatComYaoWang2013}. This turns MoS$_2$ from a simple semiconductor to a material with a strong-spin orbit coupling (SOC) effect~\cite{XiaoPRL2012,shin2019,NatComYaoWang2013}.

In this paper, we use HHG spectroscopy to explore the effects of interlayer coupling in bilayer MoS$_2$ and to investigate how the pseudospin and valley-index can differentiate the monolayer from the bilayer. Furthermore, we pursue to understand whether or not the Berry Curvature and Dipole Transition Matrix Elements (DTMEs) have an impact on the harmonic emissions~\cite{XiaoPRL2012}.

\noindent Pioneering studies of HHG spectra from monolayer and bilayer graphene have shown qualitative differences~\cite{Yoshikawa2017,Dixit2021} and, in addition, shown an attractive un-typical enhancement of the emissivity as a function of the ellipticity of the driven laser, particularly when the fundamental field is elliptically polarized. Similar effects have also been theoretically observed by Tancogne-Dejean in bulk MgO~\cite{NicolasNatComm2017} by the fully time-dependent density functional theory~\cite{AngelOctopus2020}.  

\noindent Here, we calculate the HHG spectra, using the time-dependent density matrix and analyze them using a tight-binding model (TBM), described in Sections~\ref{sec:CTDDM0}~and~\ref{MonoAndBiLayerHam0}. The Berry curvature around the K/K$'$ under the $\bm{k}\cdot \bm{p}$ approach yields~\cite{XiaoPRL2012,NatComYaoWang2013,AndorPRB2018,Wei2012}:
\begin{eqnarray}
\Omega_v({\bf k}) = -\tau\frac{2 a_0^2\,t_1^2\,M_0}{\left[ M_0^2 + 4 (a_0 t_1 k)^2 \right]^{3/2}} \label{eqn:kpBCurva01}
\end{eqnarray}
The parameters in Eq.~(\ref{eqn:kpBCurva01}) are described in Ref.~\cite{NatComYaoWang2013} for the monolayer of TMDC. Note, $M_0$ is related to the onsite potential of the lattice~(Fig.~\ref{fig:figure1}(a)), $t_1$ is nearest neighbor (NN), and $a_0$ is lattice constant and $\tau$ the valley index for this simple monolayer model.
this is the Berry curvature for the monolayer system MoS$_2$ of our TBM up to a linear expansion of the field-free Hamiltonian ${\hat H}_0({\bf k})$ obtained from Refs.~\cite{AlexisPRB2020,DenitsaPRA2021}.
According to Korm\'anyos {\it et al.} in Ref.~\cite{AndorPRB2018} the bilayer Berry curvature about the K/K', for the valence band, is written as
\begin{eqnarray}
\Omega_v({\bf k}) = \frac{\tau}{2 \delta E_{ll}}\left[ \lambda_1 {\pm} \frac{\lambda_2}{\left(1 + (\lambda_3 k/\delta E_{bg})^2\right)^{1/2}} \right] \label{eqn:kpBCurva02}
\end{eqnarray}
\noindent The parameters $\tau$, $ \delta E_{ll}$, are defined according to the spin-valley index and the onsite energies~Ref.~\cite{AndorPRB2018}. $\lambda_1$ and $\lambda_2$ are related to the hoping strength between the layers of MoS$_2$. Interestingly the bilayer Berry curvatures shows dependence of the interlayer interaction strength. Suggesting the main hypothesis of this paper: ``whether or not HHG can encode information of the monolayer peudospin and valley indexes once the system is subjected to strong middle infrared (MIR) lasers.''

\noindent We numerically examine the HHG spectra in terms of the Berry Curvature and the DTMEs for both monolayer and bilayer MoS$_2$. The results show an interesting interconnection between the selection rules as well as a particular enhancement in the HHG spectrum for a few harmonic orders (HOs). We define HO as the ratio between the emitted photon frequency $\omega$ and the freq. of fundamental driver $\omega_0$, i.e., ${\rm HO}=\omega/\omega_0$.\\ 
\noindent Further, The analysis of angular rotation, pattern of emission reveals a slight difference between mono and bilayers for the high-order harmonics. However, for the harmonic about the bandgap (Fig.~\ref{fig:figure6}), we find an interesting difference which might be attributable to the Berry Curvature and the DTMEs. These quantities contain pseudospin and valley indexes information of Mo$_2$ for monolayer or bilayer. This suggest that the angular rotation of the harmonics can be linked to DTMEs.
\noindent Surprisingly, in our HHG produced by RCP light, we notice that the HOs in the plateau region exhibit a layer difference with respect to the Berry Curvature and DTMEs. We take advantage of these quantities to analyze our numerical results, the selection rules that these quantities impose and the accumulating phase of the electron and hole wavepackets at each bands (Appendixes~\ref{sec:KeldyshSFA}). 
In the next, we will review the total currents on the Bloch basis and split them into interband and intraband currents. 
Additionally, we will describe the time-dependent density matrix formalism to compute the density at the time and ${\bf k}$--space.

%%%%%%%%%%%%%%%%%%%%%%%%%%%%%%%%%%%%%%%%%%
\section{Current and Time-dependent Density Matrix}\label{sec:CTDDM0}
\noindent The microscopic electron-charge current ${\bm j}({\bf k}, t)$ in a periodical crystalline structure subjected to an external oscillating laser ${\bf E}(t)$, and the macroscopic `measurement' is calculated by integrating the ${\bf k}$-elementary-microscopic currents in the BZ (atomic units are used throughout this paper unless otherwise indicated): 
\begin{eqnarray}
    {\bf J}(t) = \int_{\text{BZ}} \frac{d^{2}k}{(2\pi)^{2}} \bm{j}({\bf k}, t). \label{eq:current0}
\end{eqnarray}
\noindent Here we define the elementary-microscopic current ${\bm j}({\bf  k},t)$ as,
\begin{eqnarray}
    \bm{j}({\bf k}, t) &=& e\,\text{Tr}\left({\hat\rho} \hat{\bm{v}}\right) \nonumber\\
    &=& e\,\text{Tr}\left({\hat\rho}({\bf k}, t) {\hat{\bm p}}\right) \nonumber\\
    &=& e\sum_{m,n}\rho_{mn}({\bf k}, t){\bf P}_{nm}({\bf k}) \label{eq:current1}
\end{eqnarray}
This is the expectation value of the {\it velocity operator} ${\hat{\bm v}}=-{\mathrm i}\left[{\hat H}(t),\hat{\bm x} \right]$ where $ \hat{H}(t) $ is the full time-dependent Hamiltonian in the length-gauge ${\hat H}(t) = {\hat H}_0+{\hat {\bm x}}\cdot {\bf E}(t)$.~The current $ \bm{j}({\bf k}, t)$ is defined in terms of the density matrix $\hat{\rho}=\hat{\rho}({\bf k}, t)$, and the momentum matrix element ${\bf P}_{m,n}({\bf k}) =  \langle u_{m,{\bf k}} | \partial_{\bf k} {\hat H}_0({\bf k}) | u_{n,{\bf k}}\rangle$~\cite{YueGaarde2020,VladPRBVGLG02013,VladPRBVGLG1,AlexisPRB2020}. Eq.~(\ref{eq:current0}) is conventionally split into two different contributions, the interband current, ${\bf J}_{\rm er}(t)$, (associated with the momentum or dipole matrix element transition between the valence and conduction bands) and the intraband current, ${\bf J}_{\rm ra}(t)$, which show how the electron-wave and hole-wave move in each band (See Fig.~\ref{fig:figure1}(c) and~\ref{fig:figure1}(d), respectively). Mathematically, those contributions are written as follows~\cite{YueGaarde2020,AlexisPRB2020,DenitsaPRA2021}:  
\begin{eqnarray}
{\bf J}_{\rm er}(t) &=&  e\displaystyle{\sum_{ \substack{m,n \\ m \neq n}}}\int_{\text{BZ}} \frac{d^{2}k}{(2\pi)^{2}}\rho_{mn}({\bf k}, t){\bf P}_{nm}({\bf k}), \label{eq:intercurrent} \\
{\bf J}_{\rm ra}(t)&=&e\displaystyle{\sum_{ \substack{m,n \\ m = n}}}\int_{\text{BZ}} \frac{d^{2}k}{(2\pi)^{2}}\rho_{mn}({\bf k}, t){\bf P}_{nm}({\bf k}) \label{eq:intracurrent}
\end{eqnarray}
\noindent which obviously lead to the total current, 
\begin{eqnarray}
    {\bf J}(t) &=& {\bf J}_{\rm er}(t) + {\bf J}_{\rm ra}(t). \label{eq:current2}
\end{eqnarray}
To further evaluate the time-propagation density matrix ${\hat \rho}({\bf k},t)$, and the interband and intraband currents,  we numerically solve, in the time--{\bf k} space, the density matrix given by the Liouville-von Neumann equation,
\begin{eqnarray}
    i\frac{\partial {\hat \rho({\bf k}, t)}}{\partial t} = \left[ \hat{H}({\bf k},t), {\hat \rho({\bf k}, t)} \right] - i\left[ \hat{D}_2, {\hat \rho({\bf k}, t)} \right], \label{eq:Liouville1}
\end{eqnarray}
\noindent where ${\hat \rho}({\bf k}, t)$ is evaluated in the electro-magnetic length-gauge representation of ${\hat H}(t)$. We will also consider the effects of the dissipation of scattering electrons in the lattice and potential thermal bath influences by ${\hat D}_2$. As a state-of-the-art method in high-order harmonics, the phenomenological dephasing time, $T_2$, is included in the off-diagonal of the `dissipation' ${\hat D}_2$.  We adopt our numerical method, following Refs.~\cite{SilvaPRB2019,Dasol2021theory,DenitsaPRA2021} in the moving frame of ${\overline {\rm BZ}} = {\bf K} - {\bf A}(t)$, where ${\bf K}$ and ${\bf A}(t)$ denote the crystal canonical momentum and the vector potential of the laser-field, respectively. The vector potential is computed according to ${\bf E}(t) = - \partial_t {\bf A}(t)$~\cite{VampaJPB2017,VampaPRL2014}.
\noindent Given $\hat{\rho}({\bf k},t)$ matrix and the transition `canonical' momentum matrix element, ${\bf P}_{nm}({\bf k})$, we compute the HHG spectra by Fourier transforming the total current (Eq.~(\ref{eq:current2})) and taking its absolute square. This procedure allows us to compare the nonlinear optical response from monolayer with that from bilayer MoS$_2$.
%%%%%%%%%%%%%%%%%%%%%%%%%%%%%%%%%%%%%%%%%%%%%%%%%%
\section{Hamiltonian model for monolayer and bilayer MoS$_2$}\label{MonoAndBiLayerHam0}
According to the tight-binding model (TBM), the laser-free field Hamiltonian ${\hat H}_0$ of the monolayer~\cite{XiaoPRL2012} and bilayer of TMDC, under the staking 3R-polytype symmetry for the bilayer~\cite{AndorPRB2018,LisiIOP2018,RouChemC2017} (Figs.~\ref{fig:figure1}(a) and~\ref{fig:figure1}(b)) is written as:
\begin{eqnarray}
    {\hat H}_0 &= \displaystyle{\sum_{n=1}^{N}} \left( \displaystyle{\sum_i} \epsilon_n {\hat c}_{in}^{\dagger}{\hat c}_{in} -t_1 \displaystyle{\sum_{\langle i,j\rangle}}  {\hat c}_{in}^{\dagger}{\hat c}_{jn} +\displaystyle{\sum_{\langle \langle i,j\rangle \rangle}} t_{ij}{\hat c}_{in}^{\dagger}{\hat c}_{jn}\right) + \nonumber \\ 
&\displaystyle{\sum_{n=2}^{N}} \left( t_{11}{\hat c}_{in}^{\dagger}{\hat c}_{jn-1} + h.c.\right)    \label{eq:freeham0}
\end{eqnarray}
\noindent where summatory indices $n$, $\langle i,j\rangle$, and $\langle\langle i,j \rangle \rangle$ are number of layers, the intra-layer nearest-neighbour (NN) and the next nearest-neighbour (NNN) indices. $\epsilon_n$ is the onside potential of each atom in the lattice. ${\hat c}_{jn}^{\dagger}$ And ${\hat c}_{in}$ are the creation operator at the $j^{\rm th}$ lattice site and the annihilation operator at the $i^{\rm th}$ atom of the $n^{\rm th}$ layer (Figs.~\ref{fig:figure1}(a) and~\ref{fig:figure1}(b), respectively). $t_1$, $t_2$ and $t_{11}$ are the hoping parameters of the NN, NNN intra-layers, as well as the inter-layer, respectively~\cite{AndorPRB2018,XiaoPRL2012,LisiIOP2018}. 
%%%%%%%%%%%%%%%%%%%%%%%%%%%%%%%%%%%%%%%%%%
\section{Harmonics from mono-layers and bi-layers}\label{sec:Results}
In the following sub-sections, we will demonstrate the high-order harmonics emitted from monolayer and bilayer MoS$_2$ subjected to a strong MIR laser-field. In our simulations, we use both a linearly polarized laser (LPL) and a circularly polarized laser-field (CPL). We also analyze the emitted HHG spectra  as a function of the interlayer strength $t_{11}$. 

\subsection{Response to linear polarized lasers}\label{sec:Result1}
\begin{figure}[!ht]
\begin{center}
\includegraphics[width=14cm]{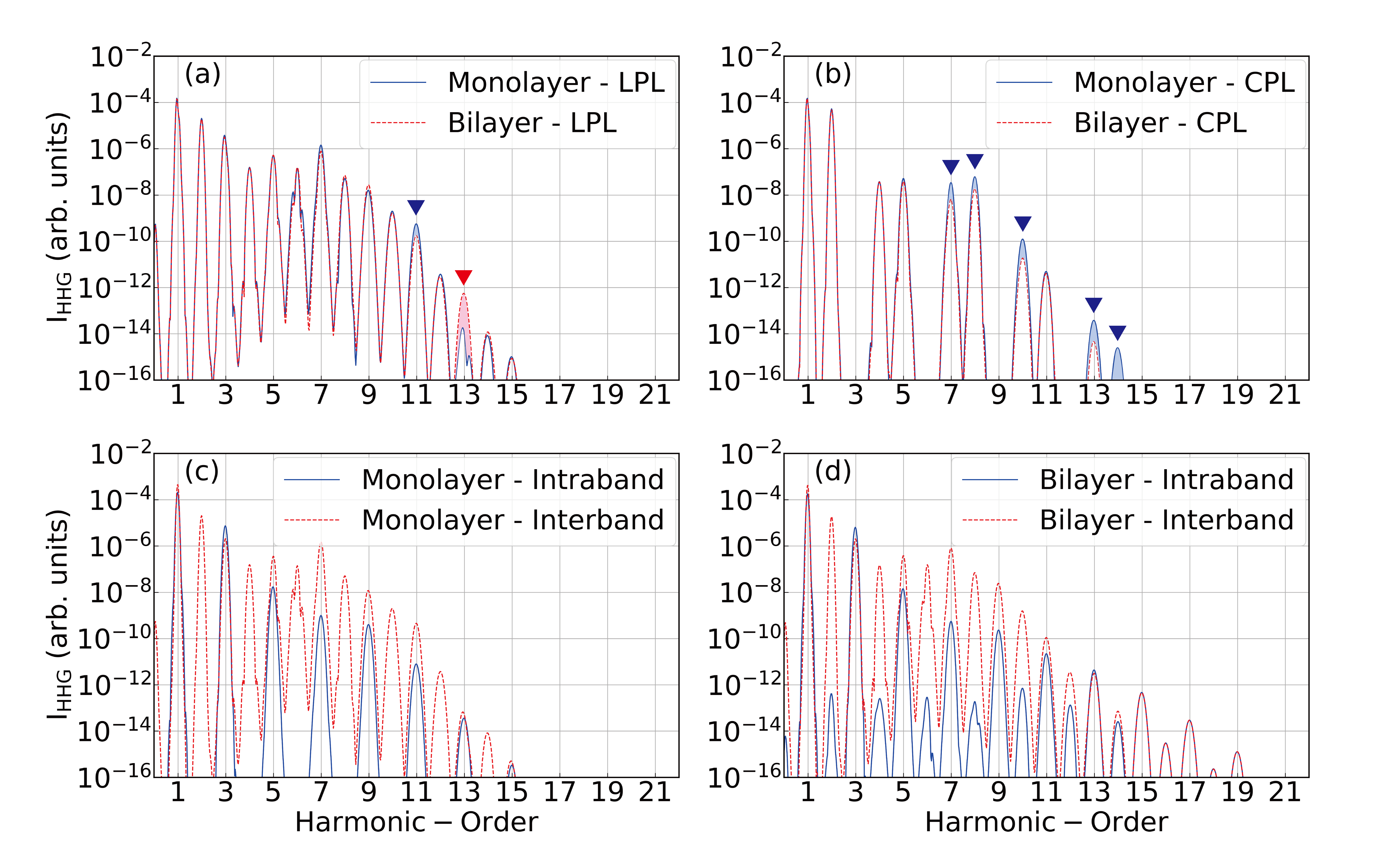}
\caption{%
 \color{black} (a) HHG spectra from monolayer (blue) and 3R phase bilayer (red) MoS$_2$ driven by linearly polarized laser along $\Gamma$-K direction. The HHG intensity is normalized with respect to the number of layers. (b) HHG spectra with the right handed circularly polarized driving pulse for monolayer (blue) and bilayer (red). (c) and (d) Interband and intraband components of HHG spectrum for monolayer and bilayer MoS$_2$, respectively, driven by linearly polarized pulse. The laser parameter used for these simulations are: $\lambda_0=3.8~\mu{\rm m}$ ($\hbar\omega_0=0.3263~{\rm eV}$), the carrier-envelop phase is fixed to zero and the number of optical cycles $N_{\rm cy}=10$ at FWHM under a gaussian envelope. The intensity of the laser is $I_0=0.315\times10^{12}~{\rm W}\cdot{\rm cm^{-2}}$, The dephasing time is fixed to $T_2=5.2$~{\it fs}. In the spectra the harmonic order (HO) corresponding to the band gap are localized about HO$6^{\rm th}$. According to our formalism, the {\it e} ({\it h}) can be driven in several region of the energy dispersion of the BZ for the monolayer and bilayer MoS$_2$ (See Figure~\ref{fig:figure6}).}
\label{fig:figure2}
\end{center}
\end{figure}

We begin presenting our numerical results by showing simulated HHG spectra from {\it monolayer} and {\it bilayer MoS$_2$} driven by a linearly (or circularly) polarized laser-fields (LPL or CPL) with respect to the $\Gamma$--K direction. The results are shown in Figs.~\ref{fig:figure2}(a) and \ref{fig:figure2}(b). 

\noindent The HHG spectra from monolayer MoS$_2$ shows a few differences with respect to the normalized (normalization of the current with respect to the number of layers) HHG from bilayer MoS$_2$. In particular, there is no substantial difference in the intensity of HHG emissions below the band gap (harmonic orders (${\rm HOs}) \leq \epsilon_0/\omega_0$). This observation is indicated in blue and red triangles. This result is surprising, since one can intuitively expect that the bilayer will enhance the emission due to more channels for tunneling excitation and interband transition.

\noindent By contrast, only the 13$^{\rm th}$ order shows an enhancement of more than one  order of magnitude for the MoS$_2$ bilayer. This can be interpreted to indicates a complex e-h recombination. In addition, Figure~\ref{fig:figure2}(a) shows the same cut-off regions for both monolayer and bilayer geometrical configurations.

\noindent The `chaotic' intensity behavior observed in Fig.~\ref{fig:figure2}(a) is rationalized via the selection rules and accumulated phase of the radiation emission, and of course, via the breaking inversion symmetry of the Hamiltonian. These selection rules are imposed by the DTMEs, ${\bf d}_{mn}({\bf k})=\,{\mathrm i}\,\langle u_{m,{\bf k}}| \partial_{\bf k}|u_{n,{\bf k}}\rangle$, and the quasi-classical action phase, $S({\bf k},t',t)$ (Appendix~\ref{sec:KeldyshSFA}), between conduction and valence bands for the monolayer system with respect to the bilayer system~\cite{YueGaarde2020,AlexisPRB2020,DenitsaPRA2021}. We think that the different coherences in the density matrix ${\hat \rho}({\bf k},t)$ lead to larger effects on the 13$^{\rm th}$-order, from the interband and intraband contributions take place (Figs.~\ref{fig:figure2}(c) and~\ref{fig:figure2}(d))~\cite{VampaJPB2017,VampaPRL2014}. While the electron (hole) wavepacket is driven in the conduction (valence) bands and its corresponding dipolar selection rules shown in Fig.~\ref{fig:figure7} and Fig.~\ref{fig:figure8} (Appendix), these ``complex propagation" effects can easily lead to constructive or destructive interfering channels. \\

\noindent These results exhibit even and odd HOs and illustrate the selection rules because the interband currents are governed by the dipole matrix element~\cite{Dasol2021theory} which breaks the inversion symmetry of the laser-field free Hamiltonian, ${\hat H}_0({\bf k})$. In case of the intraband current, the HHG spectra show a complete symmetry behaviour in the sense that only odd HOs show up in the spectra.

\subsection{Response to circularly polarized lasers}\label{sec:Result2}
\noindent The HHG produced by a contra-wise clock of the circularly polarized laser are shown in Fig.~\ref{fig:figure2}(b). For the different geometrical-phases, i.e.,~monolayer and bilayer, HHG spectra exhibit interesting differences. 
\begin{enumerate}
\item A surprising enhancement by almost one order of magnitude in the harmonics emitted around the plateau region is observed between monolayer and bilayer (See blue triangles),
\item All the harmonics follows the 3-fold symmetry of the system with the co-rotating ($3n+1$) and contra-rotating ($3n+2$) harmonic orders in the plateau region, as has already been experimentally observed in Ref.~\cite{Saito2017} in solids with a 3-fold rotational symmetry. Our results fully capture these selection rules.
As such, it is clear that circularly-polarized drivers have significant potential in bringing out the signatures between the monolayers and bilayers configurations, their differences and in particular, the valley index and pseudo-spin information as a function of the layers~\cite{AndorPRB2018}.
\end{enumerate}
\noindent As the selection rules in the HHG spectra are well described for the LPL and CPL, inversion symmetry breaking is manifested in the HHG spectra. This implies that the Berry curvature and the transition dipoles have an interesting effects on harmonic emission~\cite{GordonHandBook2006,DenitsaPRA2021,Luu2018} for both monolayer and bilayer MoS$_2$. The results shown in Fig.~\ref{fig:figure2}(b) can be attributed to Berry curvature effects about the K'-point in the BZ and the 'rotation of DTMEs' coupled to the CPL and the action phase $S({\bf k},t',t)$ (Appendix~\ref{sec:AInterBand}). Since the dipolar selection rules ${\bf P}_{mn}^{(\pm)}({\bf k}) = P_{mn}^{(x)}({\bf k})\pm iP_{mn}^{(y)}({\bf k})$ also govern the photon emission via ${\bf P}_{mn}^{(\pm)}({\bf k})$ (the momentum matrix element is proportional to the dipole transition matrix element ${\bf d}_{mn}({\bf k}) $, obviously the indexes $m\neq n$~\cite{Dasol2021theory,DenitsaPRA2021}), the RCP or LCP will excite the {\it e}-{\it h} about the K' and K point respectively. Note, however, that RCP-light will only excite electrons around the K' points of the BZ as clearly indicated in Figs.~\ref{fig:figure7}-\ref{fig:figure10}. The mathematical definition of CPL selection rule weighs the Berry curvature around the excitation {\it e} ({\it h}) wave packet and the recombination of the electron and hole. Thus, we find the potential physical picture which explains approximately the enhancement of the harmonics in the plateau region in monolayer MoS$_2$ compared to the bilayer. 

\subsection{Interlayer strength in high-order harmonics}\label{sec:Result02}
\noindent Here, we discuss the role of the inter-layer parameter $t_{11}$ in the HHG spectra. 
\begin{figure*}[!ht]
\begin{center}
\includegraphics[width=18cm]{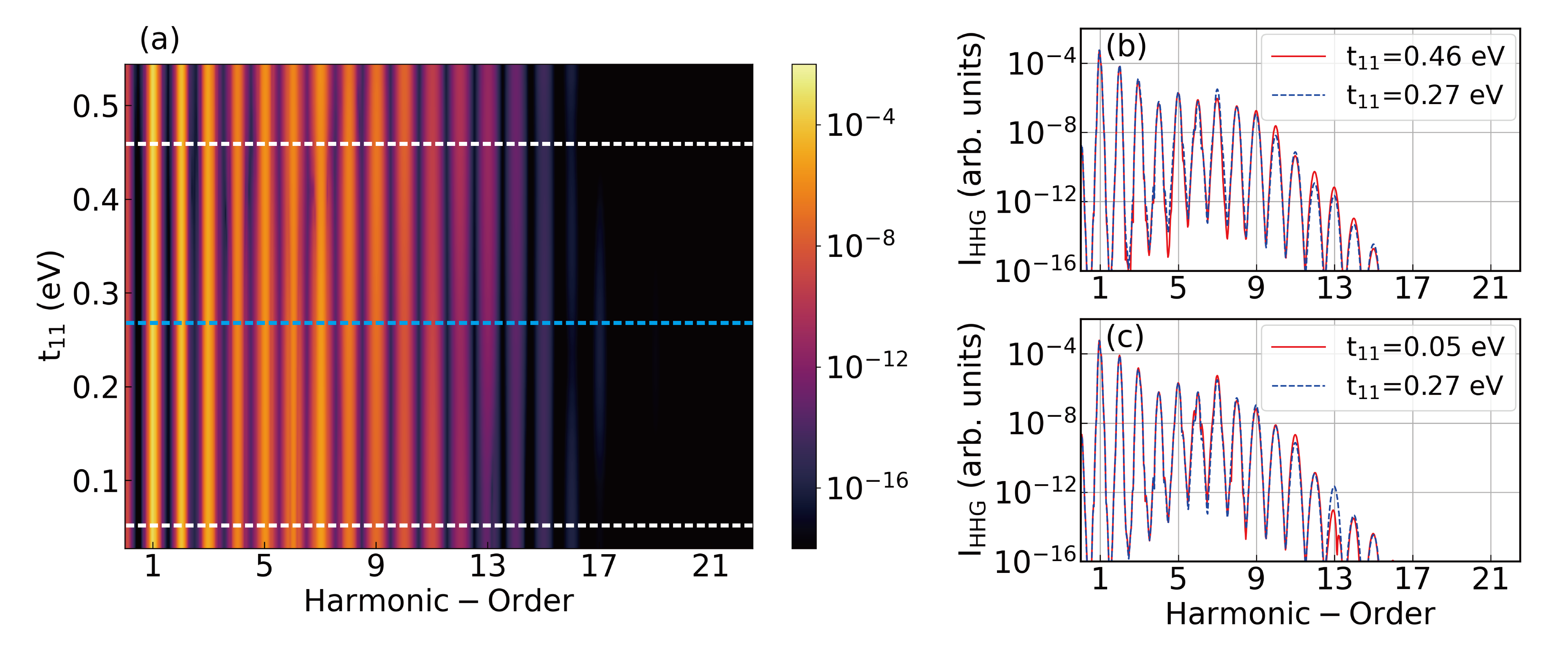}
\caption{%
(a) HHG spectra from 3R polytype bilayer MoS$_2$ as a function of interlayer coupling strength, $t_{11}$. The blue dashed line ($t_{11}$~=~0.272 eV) points out $t_{11}$ value used in most of the simulations discussed in this work. (b) and (c) show HHG spectrum for weak interlayer coupling ($t_{11}$~=~0.054 eV) and strong coupling ($t_{11}$~=~0.462 eV), respectively (white dashed lines in (a)).}
\label{fig:figure3}
\end{center}
\end{figure*}
\noindent Fig.~\ref{fig:figure3} reports the HHG spectra as a function of the inter-layer interaction $t_{11}$. We find small differences in the range of $t_{11}=\left[0,0.55\right]$~{\rm eV}. This range of $t_{11}$ is logically expected to be on the order of the magnitude of the NN parameter $t_1$. Naturally, the coupling between the atoms `Mo' of Layer1 and `S' of Layer2 is not strong enough within our approximation to observe dramatic modifications on the HHG spectrum for the bilayer MoS$_2$ (See Figs.~\ref{fig:figure3}(a), \ref{fig:figure3}(b) and \ref{fig:figure3}(c), as well as their corresponding comparisons between the approximated interaction strength of $t_{11}\sim0.27$, $0.46$ and $0.05$~eV, respectively).
However, note that once the interaction strength $t_{11}$ increases up to $2$--$3~{\rm eV}$, dramatic differences are observed between monolayer and bilayer MoS$_2$. We did not present or address this result, since it is unlikely that the interplay hopping interaction will reach such a large energy value of 2 or 3~{\rm eV}.

\subsection{Structural angular rotation of the high-order harmonics}\label{sec:Result3}
Now, We explore the angular rotation of HHG emission for low HOs (those below the band-gap), the plateau and the cut-off region as a function of the laser wavelength. 
We also study the response of the HHG spectra as a function of the ellipticity of the laser for the monolayer and bilayers of MoS$_2$. This interestingly links to the pseudo-spin and valley effects of our model into the HHG signal.
\begin{figure*}[!ht]
\begin{center}
%\begin{tabular}{cc}
%\includegraphics[width=10cm]{Fig4_2um.png}\\ 
%\includegraphics[width=10cm]{Fig4_3.8um.png} \\
\includegraphics[width=15cm]{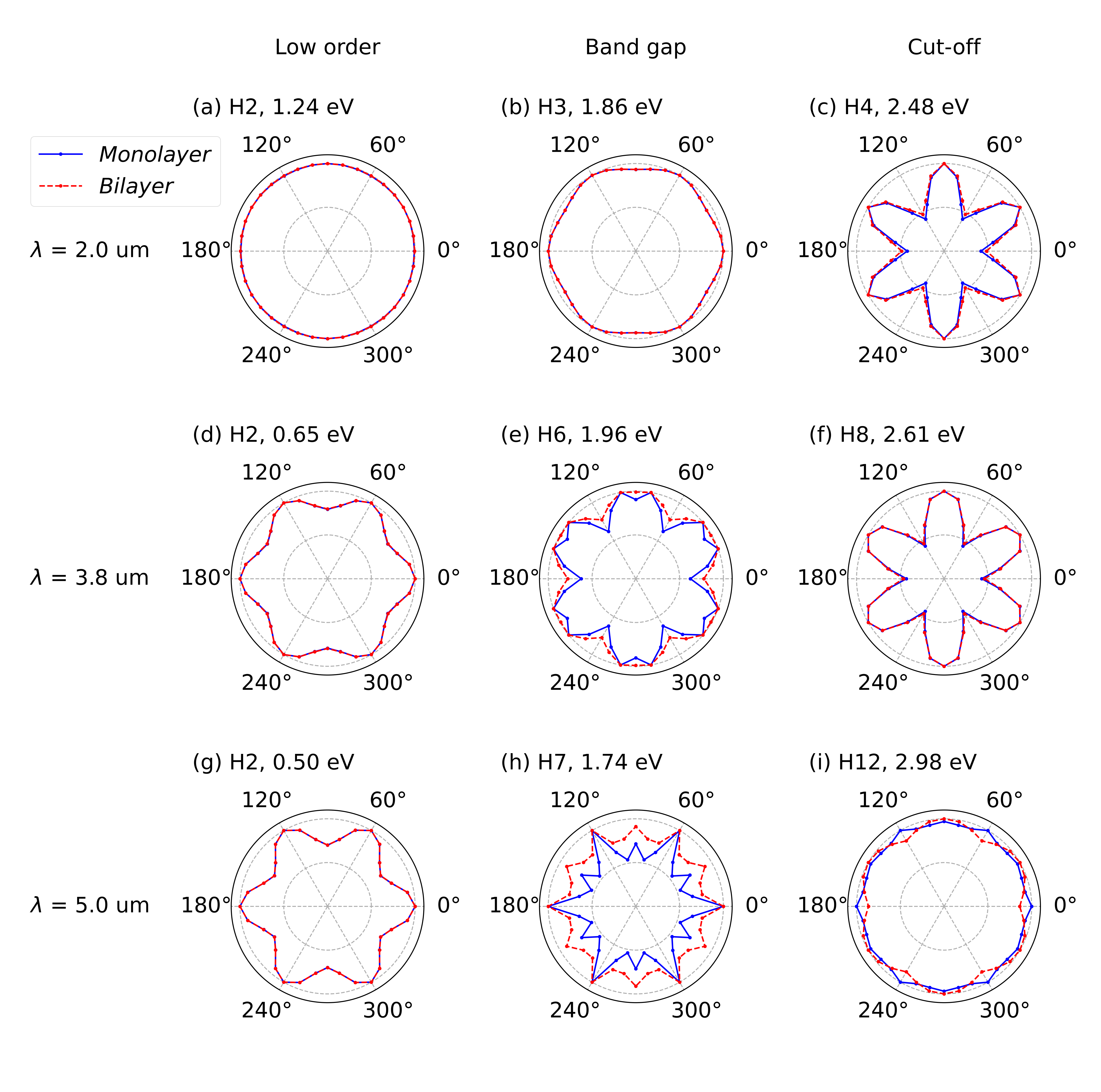}
%\end{tabular}
\caption{
Angular rotation analysis of HHG for monolayer (blue) and bilayer (red) MoS$_2$. Polarization angle dependence of harmonic yield is shown for the driving laser with the wavelength (a-c) 2~$\mu$m, (d-f) 3.8~$\mu$m and (g-i) 5.0~$\mu$m. The low order harmonics (below the band gap), the harmonics near the band gap and the harmonics around the cut-off are plotted for each driving laser wavelength.}
  \label{fig:figure4}
  \end{center}
\end{figure*}
The angular rotation of the harmonic orders is another interesting quantity which might contain information of the band energy structure, the dipole matrix element and the Berry curvature as well~\cite{LangerNature2018,W2a,Luu2018,Liu2017}. Therefore, we calculate the angular rotation of the harmonics using the following procedure: (1) the electric field, ${\bf E}(t)$, of the driving laser is linearly polarized at an angle of $\theta_0$ with respect to the $\Gamma$--K crystal orientation ($x-$direction on the $\left(k_x,\,k_y\right)$ crystal momentum space), (2) we project the resulting charge current components $J_x(t)$ and $J_y(t)$ with respect to the laser orientation. As a result, we focus our attention on the parallel $J_{\|}(\omega)$ and perpendicular $J_{\perp}(\omega)$ components with respect to the violet laser polarization~\cite{LangerNature2018}. 

We then compute the angular rotation of the harmonic orders for the parallel and perpendicular components as a function of the laser wavelengths for low HOs, bandgap and cutoff region of the HHG spectra.~The results of the total HHG, $I_{\rm HHG}(\omega,\theta_0) = |J_{\|}|^2(\omega,\theta_0)+ |J_{\perp}|^2(\omega,\theta_0)$ are shown in Fig.~\ref{fig:figure4}.\\ 
The 2$^{\rm nd}$-order shown in Figs.~\ref{fig:figure4}(a,d,g) for different laser wavelengths can be understood in terms of DTMEs and its selection rules. The DTMEs show, in Figs.~\ref{fig:figure7} and~\ref{fig:figure8}, an `start'-like structure for mono and bilayer MoS$_2$. The difference comes from the selection rules and the laser via the Bloch theorem in the BZ, i.e., ${\bf k}= {\bf k}_0 + {\bf A}(t)-{\bf A}(t')$, where, $k_0$ and $t'$ are the excitation crystal momentum and time, $t$ the emission time. ${\bf A}(t)$ is the vector potential of the laser field ${\bf E}(t)=-\partial_t {\bf A}(t)$. Additionally, we can understand the difference easily by applying perturbative theory to our formalism. Note, however, that this is out of the scope of the current paper.

The most interesting aspect of this angular analysis, in between mono and bilayer MoS$_2$, is: the angular rotation of the harmonic orders around the band gap depicted in Figs.~\ref{fig:figure4}(e)~and~\ref{fig:figure4}(h), specially for driven wavelength $\lambda_0=3.8$~and~$5.0~\mu{\rm m}$ at HO6$^{\rm th}$~and~HO7$^{\rm th}$, respectively .

\subsection{Ellipticity dependence of low- and high-order harmonics}\label{sec:Result4}

\begin{figure*}[!ht]
\begin{center}
  \includegraphics[width=15cm]{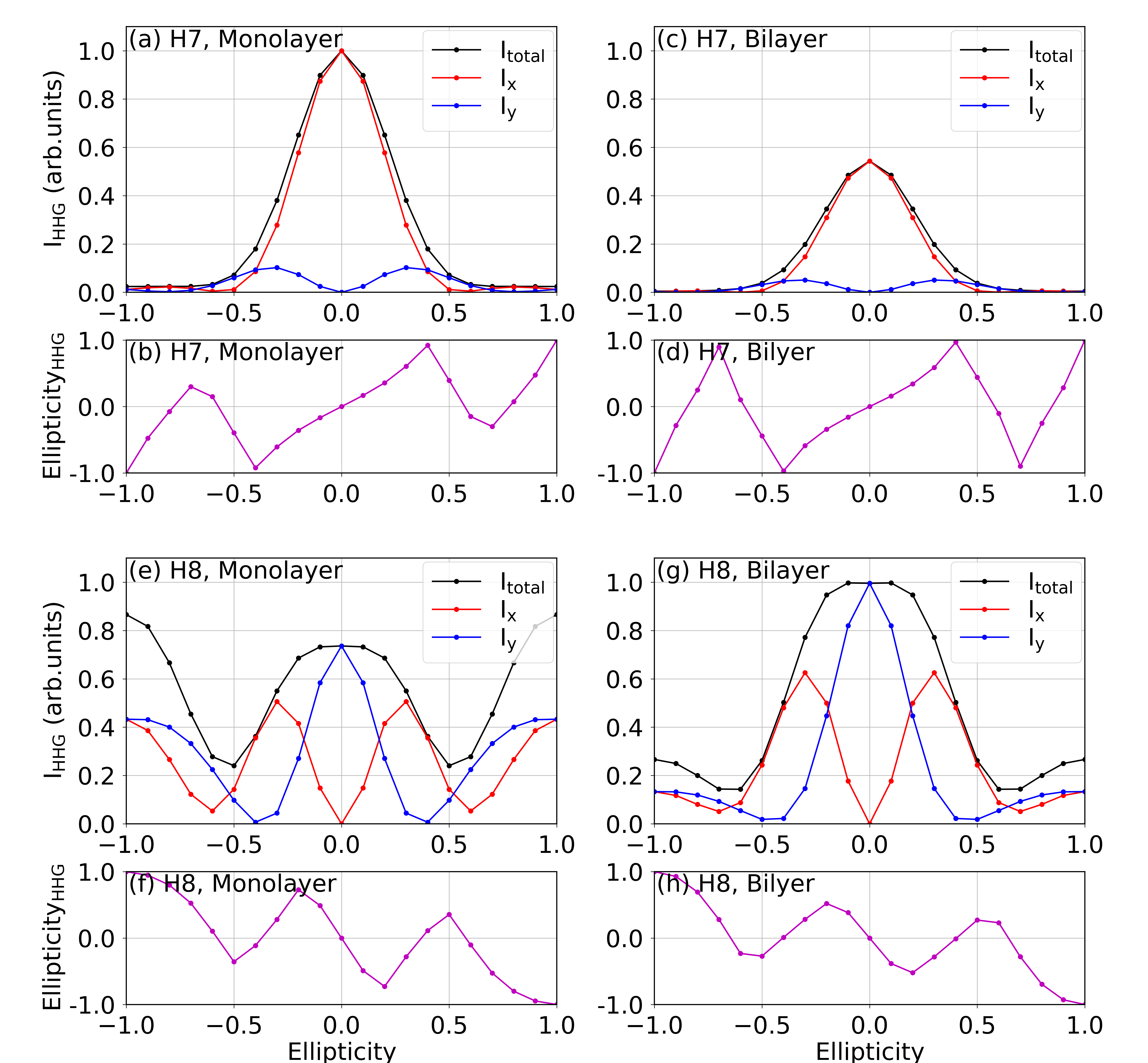}  
  \caption{%
Ellipticity profiles of the 7$^{\rm th}$ and 8$^{\rm th}$ order harmonics from monolayer and bilayer MoS$_2$. The dependence of harmonic yield on the driving laser ellipticty and the ellipticity of generated harmonic are shown (a-d) for 7th order harmonic and (e-h) for 8th order harmonic. The laser parameters are the same as those used on Fig.~\ref{fig:figure2}.
}
  \label{fig:figure5}
  \end{center}
\end{figure*}
In this section we present the result of the HHG spectra as a function of the ellipticity $\varepsilon$ of the driving-MIR laser. This $\varepsilon$ takes values between $\varepsilon=\left[-1,0,-1\right]$, meaning left-circularly polarized-laser (LCP), linear-polarized laser (LPL) and right-circularly polarized-laser (RCP). 

Figure~\ref{fig:figure5} show the calculated HHG spectra for co-rotating harmonic order (HO7$^{\rm th}$) and contra-rotating harmonic order (HO8$^{\rm th}$) between monolayer and bilayer MoS$_2$.
 Two harmonic orders, HO7$^{\rm th}$~and~ HO8$^{\rm th}$~show interesting differences as a function of the ellipticity of the incoming laser for both the monolayer and bilayer MoS$_2$. In Figs.~\ref{fig:figure5}(a,c)~and~\ref{fig:figure5}(b,d), the ellipticity of the HO7$^{\rm th}$ with respect to the ellipticity of the laser is shown. We observe almost the same tendencies with the difference that the emission yield from the monolayer is larger with respect to the bilayer emission for the LPL. The harmonic yield of the LCP vs RCP are almost zero in both monolayer and bilayer cases (Figs.~\ref{fig:figure5}(a,c)).

The HO8$^{\rm th}$, contra-rotating order with respect to the incident RCP (LCP), shows an impressive enhancement at $\varepsilon=\pm1$ (Fig.\ref{fig:figure5}~(e)) for the emission from monolayers. In contrast, the yield of HO8$^{\rm th}$ for the bilayer MoS$_2$ shows a dramatic reduction of 67$\%$ in comparison to the monolayer case (Fig.\ref{fig:figure5}~(g)).   

Finally, the contra-rotating harmonic order exhibits totally different ellipticity from the co-rotating HO. This is expected, according to the selection rules (Figs.~\ref{fig:figure9},~\ref{fig:figure10},~and~\ref{fig:figure11}). The quasi-classial action phase, $S({\bf k},t',t)$ (See Appendix~\ref{eqn:ActionML}), allows us to link this signature to pseudo-spin localized at K' or K valley. The valley index $\tau$ can be associated to the ellipticity analysis of the HOs of Fig.~\ref{fig:figure5}(f,h) with signs of the Berry curvatures for mono and bilayer MoS$_2$ at the K or K' points.
%%%%%%%%%%%%%%%%%%%%%%%%%%%%%%%%%%%%%%%%%%
\section{Conclusions and outlook}
\label{sec:conclusions}

\noindent
In summary, our theory shows that the high-harmonic generation (HHG) spectrum~is capable to:
\begin{itemize}[itemsep=0em,parsep=0.3em,topsep=0.3em,partopsep=0.3em]
\item observe difference on the nonlinear optical emission about the band gap between monolayers and bilayers of TMDCs. 
\item describe a unique difference between the angular rotations and ellipticity dependence of the emitted harmonics as a function of the number of layers concerning the ellipticity of the laser, via the selection rule of the dipole matrix element and Berry Curvature to a linear and circularly polarized laser beam.
\item is susceptible to breaking the inversion symmetries (IS, respectively) and thus sensitives to the Berry Curvature and its pseudospin character.
\end{itemize} 
\noindent These features promise the applications of the TMDC into spin-valley manipulation of the information, thanks to its geometrical and pseudo metallic properties. The TMDC in 2D can especially exhibit topological insulating phases, making them attractive for quantum computing, storage of information, and other applications~\cite{XiaoPRL2012}.
Finally, it is important to note that all of these ideas can be applied and tested in quantum simulators~\cite{AcinRoadmap}, in particular using ultracold atoms and lattice shaking~\cite{LSA12}, as well as Rydberg atoms, polaritons, or circuit QED \cite{Bloch08, specialissueNaturePhys, specialissueNaturePhys1, specialissueNaturePhys2, specialissueNaturePhys3, specialissueNaturePhys4}. 
The usage to study strong-field phenomena is an emerging application of the spintronic and valleytronic technological advances~\cite{Sala2017, Senaratne2018, Ramos2019}. Since the pseudospin is related to magnetic properties and is naturally suited to the study of systems in ordered lattices, it should provide additional clarity to the role of material with pseudospin couplings and valley indexes in high-harmonic emissions.

%%%%%%%%%%%%%%%%%%%%%%%%%%%%%%%%%%%%%%%%%%

%%%%%%%%%%%%%%%%%%%%%%%%%%%%%%%%%%%%%%%%%%
\vspace{6pt} 

%%%%%%%%%%%%%%%%%%%%%%%%%%%%%%%%%%%%%%%%%%
%% optional
%\supplementary{The following are available online at \linksupplementary{s1}, Figure S1: title, Table S1: title, Video S1: title.}


%%%%%%%%%%%%%%%%%%%%%%%%%%%%%%%%%%%%%%%%%%
\authorcontributions{Data curation and visualization, Yeon Lee; Formal analysis and investigation, Dasol Kim; Methodology and Writing-original draft, Alexis Chac\'on; Project administration and Funding Acquisition, Dong Eon Kim. Yeon Lee and Dasol Kim contributed equally to this work as first authors. All authors have read and agreed to the published version of the manuscript.}

\funding{The work has been supported in part by Grant No 2016K1A4A4A01922028 (the Max Planck POSTECH/KOREA Research Initiative Program), Grant No 2020R1A2C2103181 through the National Research Foundation of Korea (NRF) funded by the Ministry of Science and ICT, and Korea Institute for Advancement of Technology (KIAT) grant (No. P0008763, The Competency Development Program for Industry Specialists) funded by MOTIE.}

\institutionalreview{Not applicable}

%\informedconsent{}

\dataavailability{Not applicable} 

\acknowledgments{We thank the Max Planck Institute for the Structural Dynamics of matter (MPI-SD) for providing generous computational resources for the intense calculation.}

\conflictsofinterest{The authors declare no conflict of interest. The funders had no role in the design of the study; in the collection, analyses, or interpretation of data; in the writing of the manuscript, or in the decision to publish the~results.} 

%% Optional
%\sampleavailability{Samples of the compounds ... are available from the authors.}

%%%%%%%%%%%%%%%%%%%%%%%%%%%%%%%%%%%%%%%%%%
%% Only for journal Encyclopedia
%\entrylink{The Link to this entry published on the encyclopedia platform.}

%%%%%%%%%%%%%%%%%%%%%%%%%%%%%%%%%%%%%%%%%%
%% Optional
%\abbreviations{Abbreviations}{
%The following abbreviations are used in this manuscript:\\
%
%\noindent 
%\begin{tabular}{@{}ll}
%MDPI & Multidisciplinary Digital Publishing Institute\\
%DOAJ & Directory of open access journals\\
%TLA & Three letter acronym\\
%LD & Linear dichroism
%\end{tabular}}
%
%%%%%%%%%%%%%%%%%%%%%%%%%%%%%%%%%%%%%%%%%%
%% Optional
\appendixtitles{yes} % Leave argument "no" if all appendix headings stay EMPTY (then no dot is printed after "Appendix A"). If the appendix sections contain a heading then change the argument to "yes".
\appendixstart
\appendix
\section{Band gap structure in the BZ}\label{sec:AppendixA1Bands}
\begin{figure*}[!ht]
\begin{center}
\includegraphics[width=15cm]{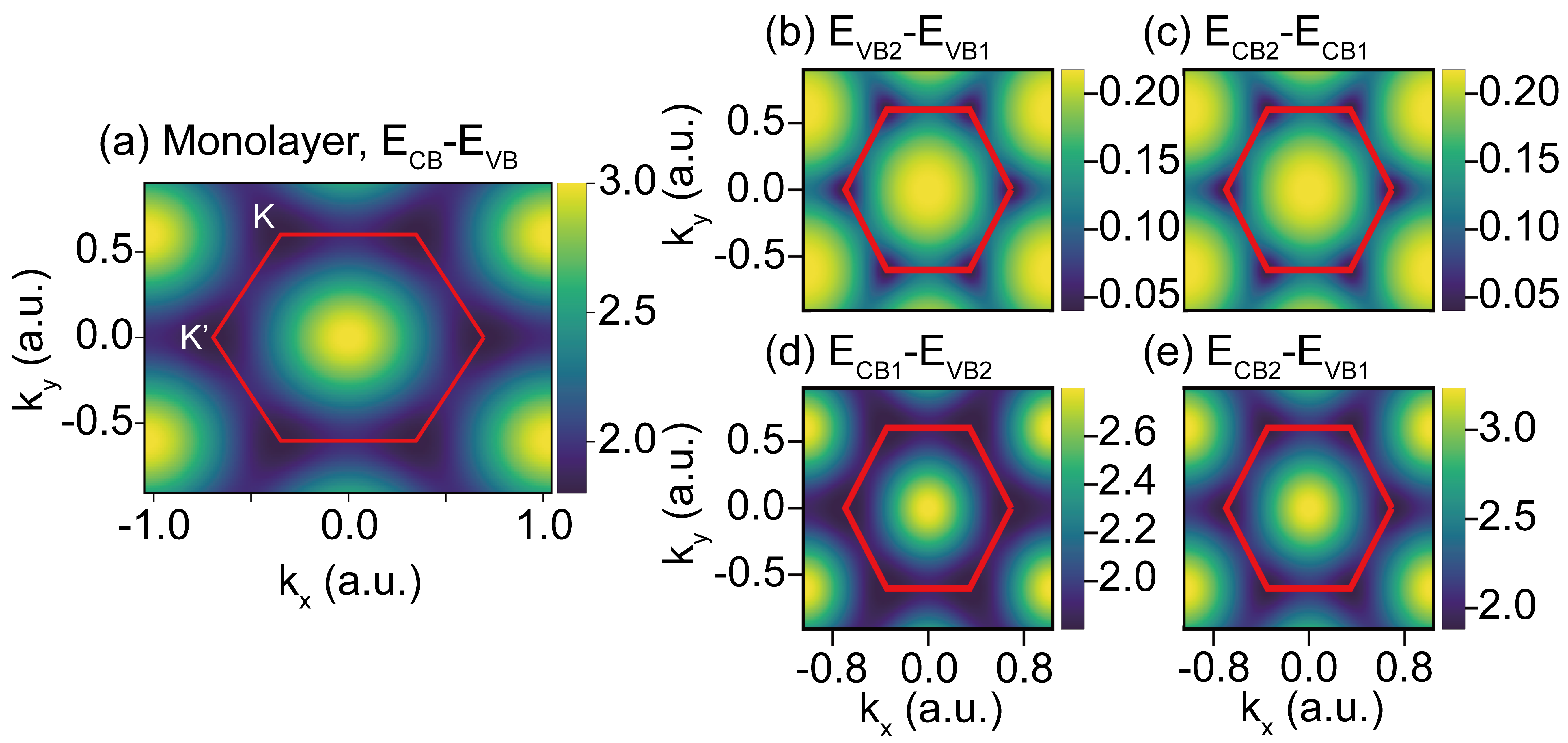}
\caption{\color{black}Deference between conduction bands (CBs) and valence bands (VBs) for monolayer and bilayers. (a) Energy difference for the monolayer model of MoS$_2$. In case of bilayer model of MoS$_2$, the energy difference (or Energy gap) for the highest VB2 and the lowest VB1 is depicted in (b), and, the energy difference of the highest CB2 and the lowest CB1 in (c), respectively. (d) And (e) depict the energy  difference for the CBs and VBs in case of bilayer MoS$_2$. The first Brillouin zone is shown as red line.}
\label{fig:figure6}
\end{center}
\end{figure*}

The energy difference band between different conduction and valence bands for monolayers and bilayers MoS$_2$ are depicted in Fig.~\ref{fig:figure6}. These energy differences are important once the high-order harmonic spectrum are dominated by the inteband currents.~\cite{VampaPRL2014,AlexisPRB2020}. We then can discover a unique physical feature by the interplay between the energy dispersion and the interband currents. 

\section{Keldysh approximation and quasi-classical analysis}
\label{sec:KeldyshSFA}
%Here $\varepsilon_{m}({\bf k})$ denotes the energy dispersion for the valence/conduction bands $m=v/c$, and $m'$ also ranges over both bands.
%We then compute the second and third terms on the right-hand side of the above equation by using Eq.~(\ref{eqn:Blount}), and we find ...
Here we can gain more physical insight into the physics of the HHG spectra by applying the so-called Keldysh approximation, as  discussed by Vampa {\it et al.}~\cite{VampaPRL2014}. This approximation reads: $\sum_m {\rho}_m({\bf k},t) \sim 0$, where $m$ is an integer index that stands for the number of conduction bands. Thus, the Keldysh approximation is valid in the limit that $a_0E_0 \sim \epsilon_0$ and $\epsilon_0 \gg \hbar\omega_0$. This essentially means that the population transferred to the conduction band is very small compared to that remaining in the valence band. This approximation is very similar to the one used in the Strong Field Approximation (SFA), which was originally developed for atoms and molecules~\cite{AlexisPRB2020,Corkum1993,Lewenstein1994,Symphony2019} -- we will thus hereafter term it SFA.  We focus on the discussion of the inter-band current in this appendix. 

\subsection{Inter-band current}\label{sec:AInterBand}
In  this approximation, we will restrict ourselves to the harmonic radiation produced by the interband current according to~Ref.~\cite{AlexisPRB2020}, We obtain a closed form of the expression for the $i$\textsuperscript{th} vectorial-component ($i=x,\,y$) within the Hamiltonian matter-gauge~\cite{SilvaPRB2019} for 2D materials,
%Dr. A. S. mod, vectorial-components => vectorial-component  
\begin{align}
J^{(i)}_\mathrm{er}(t)
&  =
-\mathrm{i}\sum_{j}{\frac{d}{dt}}
\int_{t_0}^{t} dt'
\int_{\rm  \overline{BZ}} d^2 {\bf K}\,
d^{(i)}_{cv}\left({\bf K}+ {\bf A}(t)\right)
\nonumber \\ & \qquad \quad  \times
 d^{(j)}_{cv}\left({\bf K}+ {\bf A}(t')\right) E^{(j)}(t')
\nonumber \\ & \qquad \quad \times 
e^{-\mathrm{i}S({\bf K},t,t')-(t-t')/T_2}
+\mathrm{c.c.},
\label{eqn:interML}
\end{align}
where $S({\bf K},t,t')$ is the so-called quasi-classical action for the electron-hole, which is defined as
\begin{align}
S({\bf K},t,t')
& =
\int_{t'}^{t}
\left[
  \varepsilon_g({\bf K} + {\bf A}(t'')) 
  \right. \nonumber \\ & \qquad \qquad \left.
  + \, {\bf E}(t'')\cdot{\bm \xi}_g({\bf K}+{\bf A}(t'')) 
  \right]dt'' .
\label{eqn:ActionML}
\end{align}
Here, $j=x,y$ indicates the component of the electric field and the transition-dipole product which depends on the polarization of the driving laser.
Expressions \eqref{eqn:interML} and \eqref{eqn:ActionML} are direct analogues of the Landau-Dykhne formula for HHG in atoms, which was derived in Ref.~\cite{Lewenstein1994}, following the idea of the simple man's model~\cite{Corkum1993} (See \cite{Symphony2019} for a recent review). 
Below we will analyze these expressions using the saddle point approximation over crystal momentum to derive the effects of the Berry curvature on the relevant trajectories.

\begin{figure}[!ht]
\begin{center}
\includegraphics[width=14cm]{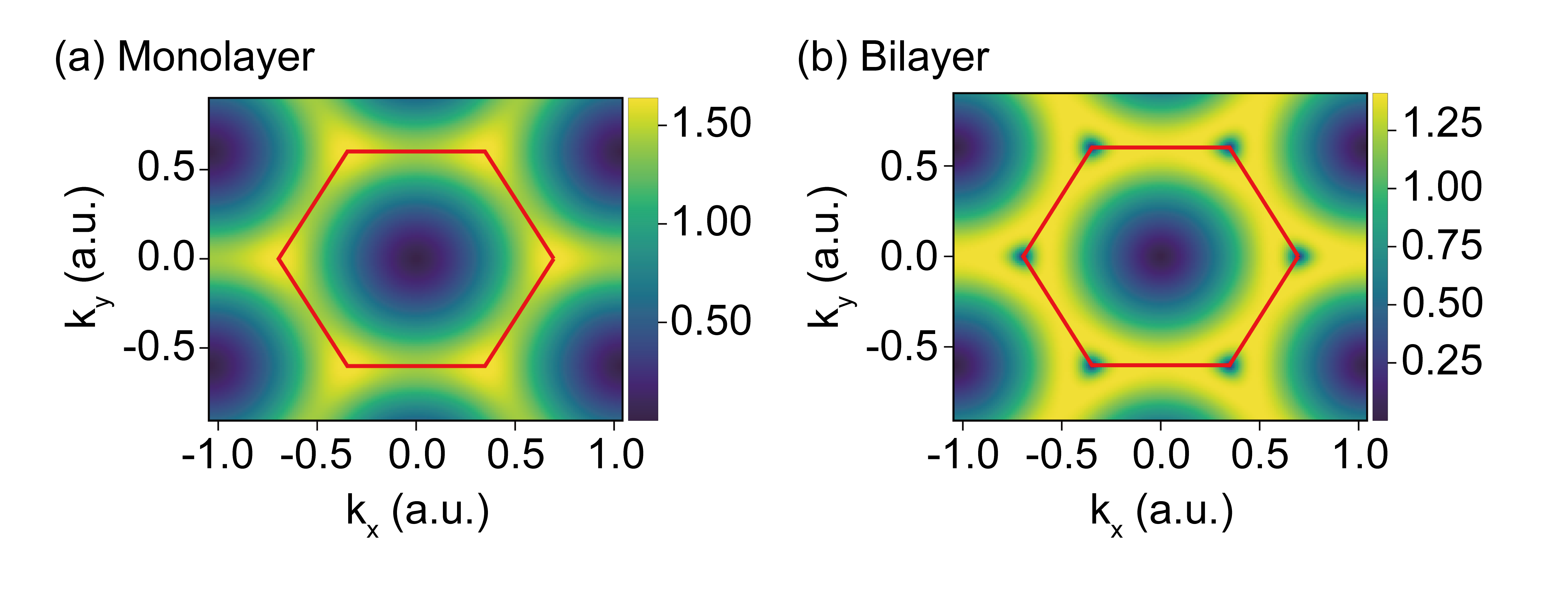}
\caption{Absolute dipole transition matrix element for (a) monolayer and (b) bilayer MoS$_2$. (a) Absolute value of dipole transition matrix elements between conduction and valence is plotted for monolayer. (b) For bilayer case, absolute value of dipole transition matrix element between highest VB2 and lowest CB1 is plotted.}
\label{fig:figure7}
\end{center}
\end{figure}

\begin{figure}[!ht]
\begin{center}
\includegraphics[width=14cm]{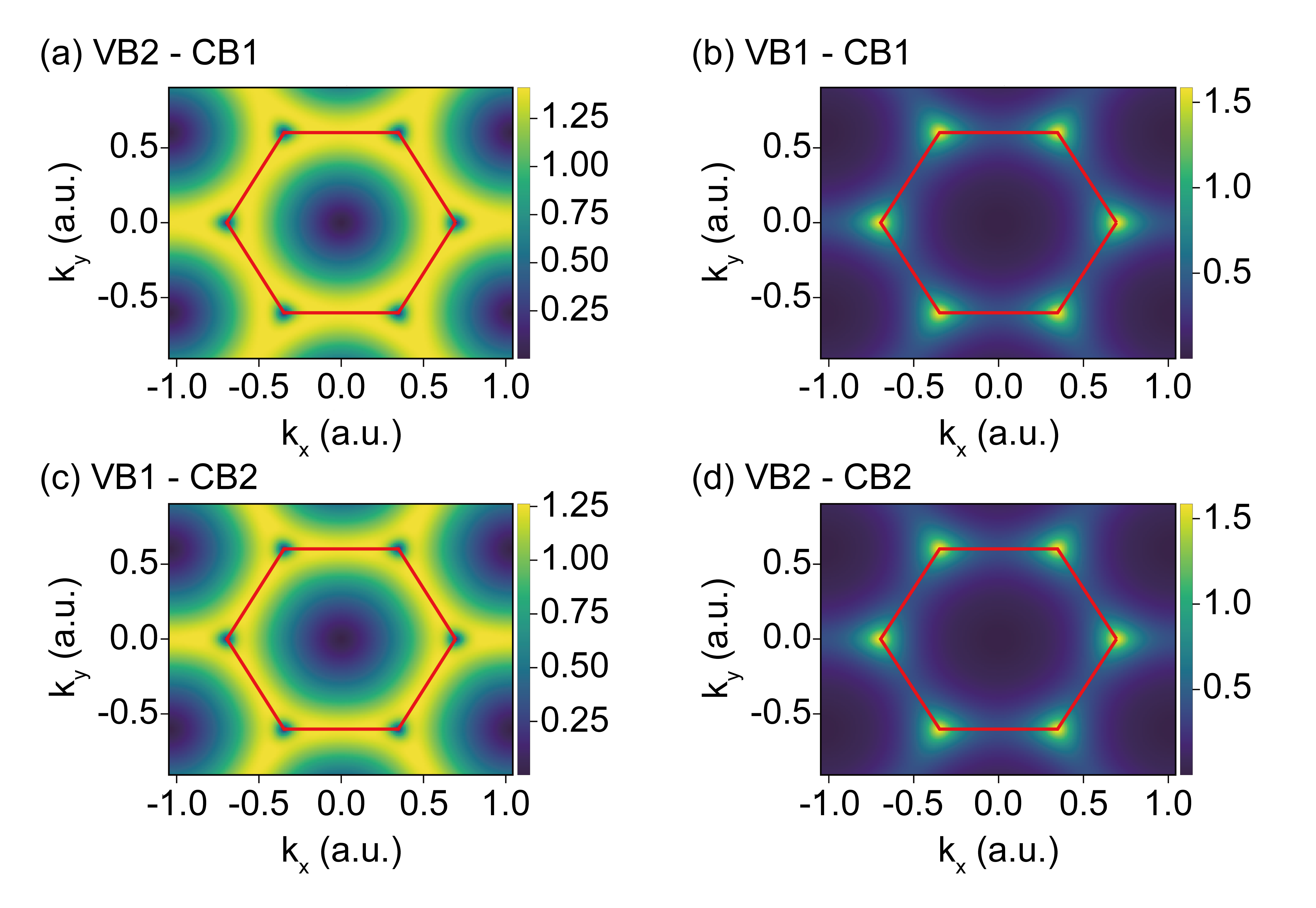}
\caption{Dipole transition matrix element between VBs and CBs for bilayer MoS$_2$. Absolute value of dipole transition matrix elements for bilayer case corresponding to (a) hightest VB2 - lowest CB1, (b) lowest VB1 - lowest CB1, (c) lowest VB1 - highest CB2, and (d) highest VB2 - highest CB2.}
\label{fig:figure8}
\end{center}
\end{figure}

\begin{figure}[!ht]
\begin{center}
\includegraphics[width=14cm]{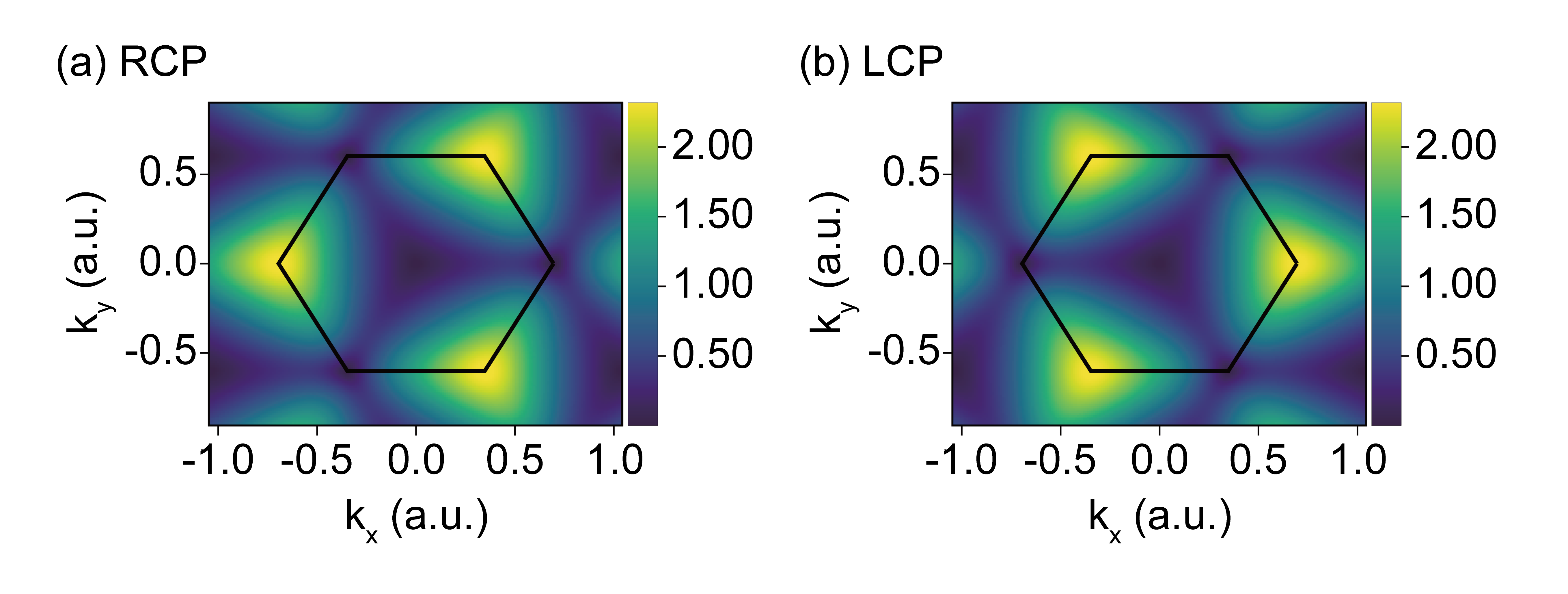}
\caption{Rotational selection rules of dipole transition matrix element for monolayer MoS$_2$ while the material interact with circularly polarized laser. Rotational dipole corresponding to (a) right (counter clockwise) circularly polarized laser and (b) left (clockwise) circularly polarized laser is plotted.}
\label{fig:figure9}
\end{center}
\end{figure}

\begin{figure}[!ht]
\begin{center}
\includegraphics[width=14cm]{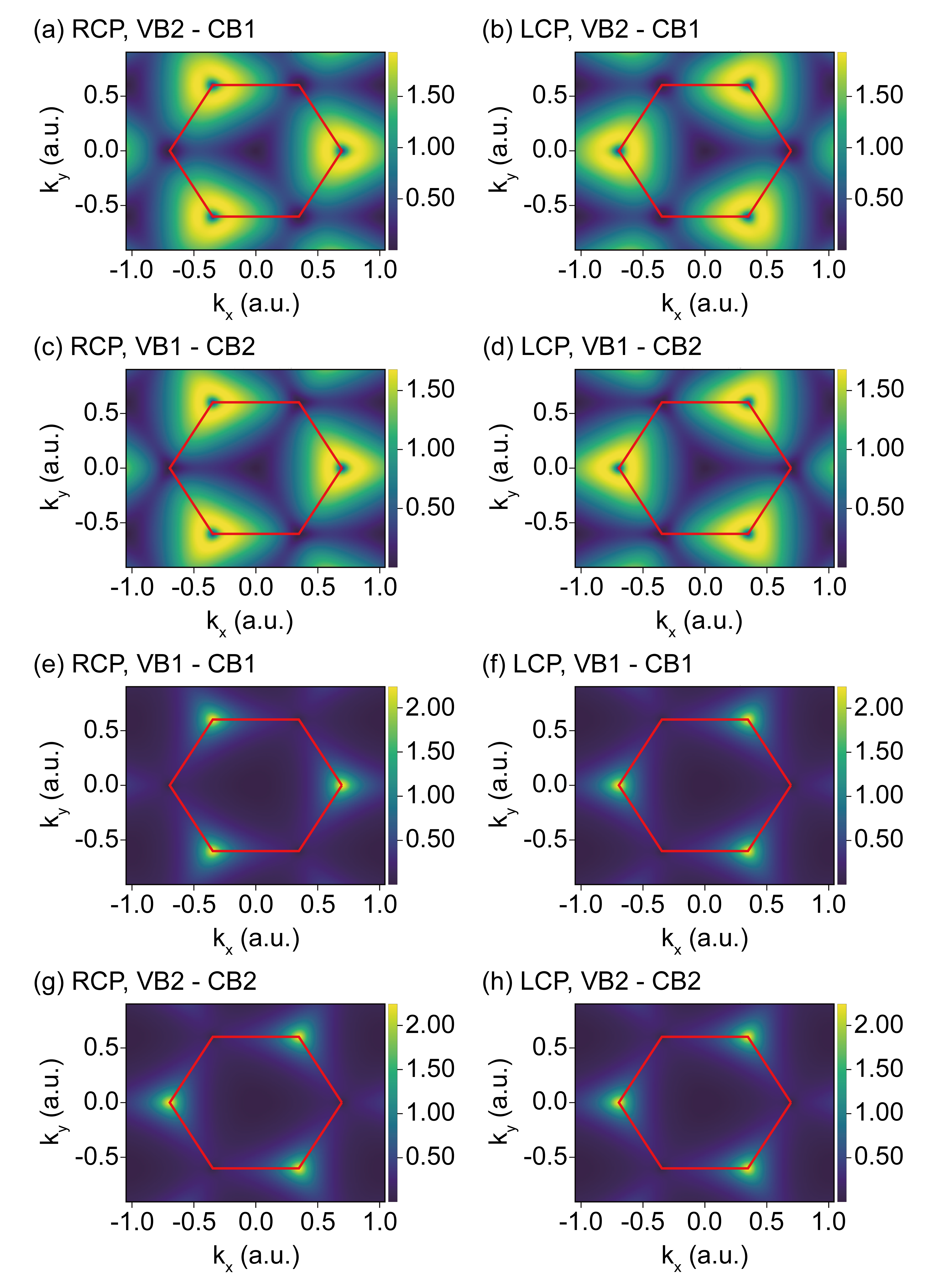}
\caption{Rotational selection rules of the dipole transition matrix element for bilayer MoS$_2$ while the material interact with circularly polarized laser. Left column (a, c, e, g) show rotational dipole for right circularly polarized laser, and right column (b, d, f, h) is for left circular polarization. Each rows are corresponding to different interband transition between VBs and CBs. Corresponding interband transitions are (a, b) highest VB2 - lowest CB1, (c, d) lowest VB1 - highest CB2, (e, f) lowest VB1 - lowest CB1, and (g, h) highest VB2 - highest CB2.}
\label{fig:figure10}
\end{center}
\end{figure}

\begin{figure}[!ht]
\begin{center}
\includegraphics[width=14cm]{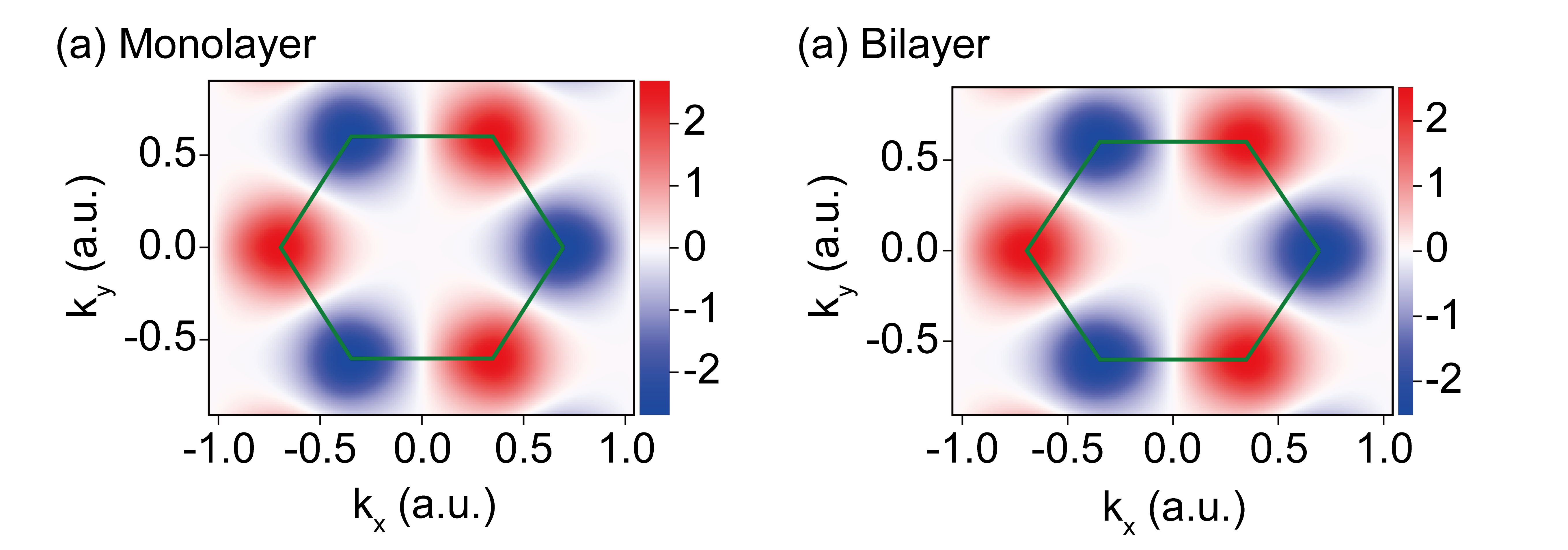}
\caption{Selection valley Berry curvature effect for monolayer and bilayer MoS$_2$. (a) And (b) show the Berry curvature of the valence band for the monolayer and bilayer, respectively. We use the average of the Berry curvature of the valence bands.}
\label{fig:figure11}
\end{center}
\end{figure}

\subsection{Quasi-classical electron-hole pair quantum paths}\label{sec:SPASBE}
Assuming that the exponentiated quasi-classical action $e^{-iS}=e^{-iS({\bf K},t,t')}$ oscillates rapidly as a function of the canonical crystal momentum $\bf K$, one can apply the saddle-point approximation to find the points ${\bf K}_s$  at which the integrand's contributions to the inter-band current (\ref{eqn:interML}) concentrated. These are solutions to the saddle-point equation $\nabla_{\bf K} S({\bf K},t,t')|_{{\bf K}_s}\approx {\bf 0}$, which can be rephrased as
\begin{equation}
\Delta{\bf x}_{c}({\bf K}_s,t,t')-\Delta{\bf x}_{v}({\bf K}_s,t,t')\approx {\bf 0}.
\label{eqn:SaddleP1}
\end{equation}
Two different trajectories are identified from the last equation, the first one is related to the excited electron $\Delta{\bf x}_{c}({\bf K}_s,t,t')$ in the conduction band, whereas the second one involves the trajectory $\Delta{\bf x}_{v}({\bf K}_s,t,t')$, followed by the hole in the valence band.
We then obtain a general $m$\textsuperscript{th} trajectory for the electron ($m=c$) and hole ($n=v$), which is expressed as
\begin{align}
\Delta{\bf x}_{m}({\bf K}_s,t,t') 
& =
\int_{t'}^{t}
\bigg[
  {\bf v}_{\mathrm{gr},m}
  + {\bf E}(t'')\times{\bm \Omega}_{m}
\label{eqn:SaddleP3}
\\ & \!\!
\nonumber 
  +\left({\bf E}(t''){\cdot}{\nabla_{\bf K}}\right)\!
  \bigg({\bm \xi}_{m}  
  {+}
  (-1)^m \frac{1}{2}\nabla_{\bf K}\phi^{(j)}_{cv}\bigg) 
  \bigg]dt''
,
\end{align}
where $(-1)^m$ is the alternating sign $(-1)^{c}=+1$ and $(-1)^v=-1$,
and the group velocity of the $m$\textsuperscript{th} band is ${\bf v}_{\mathrm{gr},m} = \nabla_{\bf K}\varepsilon_m$. Here we recognize the Berry curvature ${\bm \Omega}_m$ as well as the anomalous velocity ${\bf v}_{a,m}$, which are given by ${\bm \Omega}_m={\nabla}_{\bf K}\times{\bm\xi}_m$ and ${\bf v}_{a,m}={\bf E}(t)\times{\bm\Omega}_m$, respectively, for the electron-hole trajectories of Eq.~(\ref{eqn:SaddleP3}). The previous expression can be rewritten as
\begin{align}
\Delta{\bf x}_{m}({\bf K}_s,t,t')
& =
\int_{t'}^{t} \!
\bigg[
  {\bf v}_{\mathrm{gr},m}  
  + {\bf v}_{a,m} 
\label{eqn:SaddleETs}
\\ & \qquad 
\nonumber 
  -\frac{d}{d t''}\left({\bm \xi}_m{+}\frac{(-1)^m}{2}\nabla_{\bf K}\phi^{(j)}_{cv}\right)  
  \bigg]dt''
.
\end{align}
These electron-hole pair trajectories, along with the saddle-point condition of Eq.~(\ref{eqn:SaddleP1}), should produce com\-plex-valued solutions for ${\bf K}_s$, as is the case for HHG of gases.
However, determining the solutions ${\bf K}_s$ is not a trivial task, as it depends explicitly on the geometrical features, such as the Berry curvature and connection, and the phase of the dipole matrix elements.
This gets further complicated as the eigenstates and eigenvalues that make up the energy bands are expected to exhibit branch cuts and branch points connecting the two bands~\cite{Pechukas1976, Hwang1977}. Once the momentum is allowed to take on complex values (as it does in complex band structure theory~\cite{Reuter2016}), this leads to a nontrivial geometrical problem with a high dimensionality whose analysis requires detailed attention.

Nevertheless, these saddle points ${\bf K}_s$ should have a component perpendicular to the driving laser field ${\bf E}(t)$ (in the case of linear drivers), which appears as a consequence of anomalous-velocity features and of the Berry curvature ${\bm \Omega}_m({\bf k})$ in particular.

\subsection{Berry Connection, Berry Curvature, and Chern number}
To calculate the radiation-interaction features of these eigenstates, the DTME between each state of from conduction to valence bands or vice versa. This DMTE is unique, since it contains the full information of the selection rules for the emitted radiation. We define it in the basis of the Bloch function~\cite{Blount1962}, 
\begin{align}
\langle \Phi_{m,{\bf k}'}| {\bf x}|\Phi_{n,{\bf k}}\rangle
& =
-\mathrm{i}\nabla_{\bf k} \left( \delta_{mn}\delta({\bf k}-{\bf k}')\right) 
\nonumber \\ & \qquad \qquad 
+ \delta({\bf k}-{\bf k}'){\bf d}_{mn}({\bf k})
,
\label{eqn:Blount}
\end{align}
where
\begin{align}
{\bf d}_{mn}({\bf k}) &= \mathrm{i}\,\langle u_{m,{\bf k}}| \nabla_{\bf k}|u_{n,{\bf k}}\rangle
\label{eqn:DTME0}
\end{align}
is a regular function which encodes the momentum gradient of the periodic part of the Bloch functions.
For $n\neq m$, ${\bf d}_{mn}({\bf k})$ defines the DTME as also have been noticed in the text.
 The main diagonal of Eq.~(\ref{eqn:DTME0}) elements is the Berry connection of the $m^{\rm th}$-band,
\begin{align}
{\bm\xi}_m(\mathbf {k})& = {\bf d}_{mm}({\bf k})
\nonumber \\
& =
\mathrm{i}\langle u_{m,{\bf k}}| \nabla_{\bf k}|u_{m,{\bf k}}\rangle,
\label{eqn:BerryConnection}
\end{align}
which is responsible for the parallel transport of wave function phase around the band.
This parallel transport is measured by the Berry curvature, given by the gauge-invariant curl
\begin{equation}
{\bm \Omega}_{m}({\bf k})
= \nabla_{\bf k} \times {\bm \xi}_{m}({\bf k}) \label{eqnBCurva0}
\end{equation}
This parallel transport is observed in the quasi-classical action phase of the Eq.~(\ref{eqn:DTME0}).
\subsection{Dipole moment and Berry curvature in the Haldane model (HM)}
We now turn to the dipole transition matrix element, as initially defined in Eq.~\eqref{eqn:DTME0}, and its deep relationship with the Berry curvature which is given here as~\cite{AlexisPRB2020}
\begin{equation}
{\bf d}_{cv}({\bf k})
=
\frac{1}{2} \left[ (\sin\theta_{\bf k}) (\nabla_{\bf k}\phi_{\bf k}) + i \nabla_{\bf k}\theta_{\bf k} \right].
\label{eqn:DME1}
\end{equation}
On the one hand, we  find that the cross product of the HM dipole Eq.~\eqref{eqn:DTME0} yields
\begin{align}
\Im(d_{cv}^{x}d_{cv}^{y*})
& =
\frac{1}{4}\sin \theta_{\bf k}
\left[
    \partial_{k_y}\!\theta_{\bf k} \ \partial_{k_x}\!\phi_{\bf k}
    -\partial_{k_x}\!\theta_{{\bf k}} \ \partial_{k_y}\!\phi_{{\bf k}}
    \right].
\label{eqn:HMCrossDipole}
\end{align}
Expanding the Berry curvature in the Haldane model, given by Eq.~\eqref{eqnBCurva0}, one obtains~\cite{AlexisPRB2020,Wei2012}
\begin{align}
{\bf \Omega}_{v/c}({\bf k})
& =
\mp \frac{1}{2}\sin\theta_{\bf k}(\nabla_{\bf k}\theta_{\bf k}\times\nabla_{\bf k}\phi_{\bf k})
\\
\nonumber 
& =
\mp \hat{\bf z}\frac{1}{2}\sin\theta_{\bf k}
\left[
    \partial_{k_y}\!\theta_{\bf k} \ \partial_{k_x}\!\phi_{\bf k}
    -\partial_{k_x}\!\theta_{{\bf k}} \ \partial_{k_y}\!\phi_{{\bf k}}
    \right].
\end{align}
We therefore conclude that~\cite{AlexisPRB2020}:
\begin{equation}
\Omega_{v/c}  
= 
\mp 2\Im[d_{cv}^{(x)}d_{cv}^{*(y)}]
,
\end{equation}
which demonstrates the close relationship between dipole matrix elements and the Berry curvature in the HM. This  confirmation of the relationship between the dipole product and the Berry curvature is extremely important, as it leads to a direct connection of the inter-band transition current of Eq.~\eqref{eqn:interML} and the topological invariant for this model.

%%%%%%%%%%%%%%%%%%%%%%%%%%%%%%%%%%%%%%%%%%
\end{paracol}

\reftitle{References}


\externalbibliography{yes}
\bibliography{references}

%=====================================
% References, variant B: internal bibliography
%=====================================

% If authors have biography, please use the format below
%\section*{Short Biography of Authors}
%\bio
%{\raisebox{-0.35cm}{\includegraphics[width=3.5cm,height=5.3cm,clip,keepaspectratio]{Definitions/author1.pdf}}}
%{\textbf{Firstname Lastname} Biography of first author}
%
%\bio
%{\raisebox{-0.35cm}{\includegraphics[width=3.5cm,height=5.3cm,clip,keepaspectratio]{Definitions/author2.jpg}}}
%{\textbf{Firstname Lastname} Biography of second author}

% The following MDPI journals use author-date citation: Arts, Econometrics, Economies, Genealogy, Humanities, IJFS, JRFM, Laws, Religions, Risks, Social Sciences. For those journals, please follow the formatting guidelines on http://www.mdpi.com/authors/references
% To cite two works by the same author: \citeauthor{ref-journal-1a} (\citeyear{ref-journal-1a}, \citeyear{ref-journal-1b}). This produces: Whittaker (1967, 1975)
% To cite two works by the same author with specific pages: \citeauthor{ref-journal-3a} (\citeyear{ref-journal-3a}, p. 328; \citeyear{ref-journal-3b}, p.475). This produces: Wong (1999, p. 328; 2000, p. 475)

%%%%%%%%%%%%%%%%%%%%%%%%%%%%%%%%%%%%%%%%%%
%% for journal Sci
%\reviewreports{\\
%Reviewer 1 comments and authors’ response\\
%Reviewer 2 comments and authors’ response\\
%Reviewer 3 comments and authors’ response
%}
%%%%%%%%%%%%%%%%%%%%%%%%%%%%%%%%%%%%%%%%%%
\end{document}